\documentclass[aps,draft,twocolumn,a4paper]{revtex4}

\input epsf

\begin{document}

\title{
Class of exactly soluble models of one-dimensional spinless fermions
and its application to the Tomonaga-Luttinger Hamiltonian with nonlinear
dispersion
}
\author{A.V. Rozhkov}

\affiliation{
Institute for Theoretical and Applied Electrodynamics RAS,
ul. Izhorskaya 13/19, 125412, Moscow, Russian Federation
}
\date{\today}

\begin{abstract}
It is shown that for some special values of Hamiltonian parameters the
Tomonaga-Luttinger model with nonlinear dispersion is unitary equivalent
to the system of noninteracting fermions. For such parameter values the
density-density propagator of the Tomonaga-Luttinger Hamiltonian with
nonlinear
dispersion can be found exactly. In a generic situation the exact solution
can be used as a reference point around which a perturbative expansion in
orders of certain irrelevant operators may be constructed. 
\end{abstract}

\maketitle
\hfill
\section{Introduction}
\label{intro}

The Tomonaga-Luttinger model is the most important model of one-dimensional
interacting electron physics. The model describes two Fermi points of
right-moving (R) and left-moving (L) electrons. The electrons interact
through a two-body short-range potential:
\begin{eqnarray}
H_{\rm TL}=H_{\rm kin} + H_{\rm int},\label{H}\\
H_{\rm kin} = i v_{\rm F}\int_{-L/2}^{L/2} dx
\left( \colon\psi^\dagger_{{\rm L}}
\nabla\psi^{\vphantom{\dagger}}_{{\rm L}}\colon - \colon\psi^\dagger_{{\rm
R}}
\nabla\psi^{\vphantom{\dagger}}_{{\rm R}}\colon \right) ,\label{H_kin}\\
H_{\rm int} = \int dx dx' \hat g(x-x') \rho_{\rm L}(x) \rho_{\rm R}(x'),
\label{int}\\
\rho_{\rm L,R} = \colon\psi_{\rm L,R}^\dagger \psi_{\rm
L,R}^{\vphantom{\dagger}}\colon,
\end{eqnarray}
where function $\hat g(x)$ vanishes for $\Lambda |x| \gg 1$. The parameter
$\Lambda$ plays a role of the ultraviolet cutoff. The symbol $\colon
\ldots\colon$ stands for the normal ordering with respect to the
noninteracting ground state.

The kinetic energy (\ref{H_kin}) of the model (\ref{H}) is a linear function
of the electron momentum. It is a necessary condition for the exact
solubility of $H_{\rm TL}$: when the kinetic energy is linear one can apply
bosonization to find the excitation spectrum and the correlation functions of
the model \cite{haldane,boson}.

However, there are situations in which we must extend the model by including
the nonlinear dispersion term:
\begin{eqnarray}
H_{\rm nl} = v'_{\rm F}\int dx \left( \colon\! \nabla\psi^\dagger_{\rm L}
\nabla \psi_{\rm L}^{\vphantom{\dagger}} \colon +
\colon\! \nabla\psi^\dagger_{\rm R}
\nabla \psi_{\rm R}^{\vphantom{\dagger}} \colon \right).  \label{H2}
\end{eqnarray}
Such generalization immediately introduces serious complications: the
bosons become interacting.

Another extension of the system (\ref{H}) where the bosonization fails
is the quasi-one-dimensional
conductor: an array of one-dimensional Tomonaga-Luttinger wires coupled by
transverse hopping and interaction. Bosonization works poorly for the
transverse hopping has no simple expression in terms of the
Tomonaga-Luttinger bosons. 

Thus, some researchers began developing `fermionic' approaches to the
model. In a variety of contexts the one-dimensional electron systems were
described in terms of the composite fermionic degrees of freedom
\cite{fermions1,fermions2,fermions3}. More to the topic of the present
paper are Ref. \cite{glazman,teber,rozhkov_q1d} which we briefly discuss below.
 
In Ref. \cite{glazman,teber} the authors studied the effect of nonlinear
dispersion $v_{\rm F}'$ on the density-density propagator of the
Tomonaga-Luttinger Hamiltonian. They avoided using the bosonization and
instead worked with the bare fermionic degrees of freedom.

The quasi-one-dimensional case was investigated in Ref. \cite{rozhkov_q1d}
where the hybrid boson-fermion representation for the quasi-one-dimensional
Hamiltonian was proposed. The latter representation describes the high-energy
one-dimensional degrees of freedom in terms of the Tomonaga-Luttinger bosons
while the low-energy three-dimensional degrees of freedom as fermionic
quasiparticles.

Despite their success, both approaches are not systematic. Rather, they
were custom-made for particular tasks at hand. The systematic method to
handle the generalized problem (\ref{H})-(\ref{H2}) was
proposed by the author in Ref. \cite{rozhkov}. The main idea of the method
is to apply a certain canonical transformation to (\ref{H})-(\ref{H2}) which
kills marginal (in the renormalization group sense) interaction
$H_{\rm int}$. Such transformation 
maps the original model on the model of fermionic quasiparticles weakly
interacting  through an irrelevant 
operator. Due to the irrelevance of the quasiparticle interaction the
usual perturbation theory is applicable and the quasiparticles near the Fermi
points are free.

This approach allows one to calculate the retarded density-density
correlation function of the Tomonaga-Luttinger model with nonlinear
dispersion to the zeroth order in the quasiparticle interaction
\cite{rozhkov}.

The developments described above solve the long-standing problem of the
Tomonaga-Luttinger model with the nonlinear dispersion. The purpose of
this paper is refinements of the method: we will show that if, in addition
to the nonlinear dispersion, one supplies the Tomonaga-Luttinger
Hamiltonian with a certain type of irrelevant interactions the Hamiltonian
becomes exactly soluble. The system at and near this solubility point is
studied in this paper.

We proceed by noting that the classical formulation of the Tomonaga-Luttinger
model with nonlinear dispersion (\ref{H})-(\ref{H2}) is somewhat
inconsistent. It is trivial to check that the scaling dimension of
$H_{\rm nl}$ is equal to 3. Indeed, the field operator has dimension of 1/2,
the dimension of $\nabla$ is unity. Therefore, the dimension of the whole
operator is 3. In addition to $H_{\rm nl}$, there is one and only one
operator whose scaling dimension is also 3:
\begin{eqnarray}
H_{\rm int}' = ig' \int dx 
 \left[  \rho_{\rm R} \left( \colon \psi^\dagger_{\rm L}
(\nabla \psi^{\vphantom{\dagger}}_{\rm L}) \colon -
\colon (\nabla \psi^\dagger_{\rm L}) \psi^{\vphantom{\dagger}}_{\rm L}
\colon \right)
\right.\label{H'}\\
\left. -  \rho_{\rm L} \left( \colon \psi^\dagger_{\rm R}
(\nabla \psi^{\vphantom{\dagger}}_{\rm R}) \colon -
\colon (\nabla \psi^\dagger_{\rm R}) \psi^{\vphantom{\dagger}}_{\rm R}
\colon \right)\right]. \nonumber 
\end{eqnarray}
From the renormalization group view point if we are interested in the
effect of $H_{\rm nl}$ we must include $H_{\rm int}'$ as well.

Initially, it looks as if such modification leads to substantial
additional difficulties. Yet, despite this expectations, we will see that
the operator $H_{\rm int}'$ does not make our problem more complicated at
all. Moreover, if $g'$, $v_{\rm F}'$, $g$ and $v_{\rm F}$ satisfy some
special relation the Hamiltonian
\begin{eqnarray}
H_{\rm ex}&=&H_{\rm TL} + H_{\rm nl} + H_{\rm int}'  \\
&+&\text{(irrelevant operators with scaling dim} \geq 4 ) \nonumber
\end{eqnarray}
is equivalent to the system of {\it noninteracting} fermions.
Thus, the system becomes exactly soluble. 

The exact solubility gives us a chance to access qualities otherwise hidden
from the analytical investigation. For example, the spectrum of the
Hamiltonian and the statistics of excitations can be determined. Additionally,
the density-density propagator can be calculated as well. It is proportional
to the propagator of the free fermions.

Of course, a generic model (which we denote as $H_{\rm G}$, where `G' stands
for `generic') deviates from the exactly soluble case. However, it can
be shown that if such deviations are weak we can study the generic model
with the help of perturbation theory in orders of $(H_{\rm G} - H_{\rm
ex})$. In particular, the propagator of the
exactly soluble Hamiltonian is the zeroth order approximation to the
propagator of $H_{\rm G}$.

The paper is organized as follows. The exactly soluble model is derived
in Sect.\ref{exact}. Since the latter Section is very technical the reader
might get lost in the details. For those who are not specifically interested
in all the minutia there is Sect.\ref{intuitive} where we give concise
nontechnical overview of the presented derivations. This Section is fairly
self-contained so that the reader ready to settle for just a heuristic
argumentation may skip Sect.\ref{exact}. The situation of weak deviation from
the point of exact solubility is discussed in Sect.\ref{generic}.
Sect.\ref{observe} is dedicated to the application of Sect.\ref{generic}
ideas to calculations of observables. In Sect.\ref{compare} we compare our
approach with the results available in
the literature. Technically involved calculations are relegated to Appendix.

\section{Exactly soluble model}
\label{exact}

We start with the quadratic Hamiltonian:
\begin{eqnarray}
H_0 = H_{\rm kin} + H_{\rm nl} + \ldots = \label{reference}\\
\sum_p  \int dx \left\{ i p \tilde v_{\rm F}  \colon\! \psi^\dagger_{p}
(\nabla\psi^{\vphantom{\dagger}}_{p})\colon 
+ \tilde v'_{\rm F} \colon\! (\nabla\psi^\dagger_p)
(\nabla \psi_p^{\vphantom{\dagger}} )\colon \right\}+ \ldots,\nonumber 
\end{eqnarray}
where ellipsis stand for quadratic terms which contain more space
derivatives. We must assume their presence to avoid creating spurious Fermi
points besides those located at $q=0$. From the practical point these terms
cause us no trouble. They will be discussed in Sect.IV, together with other
terms whose scaling dimension is 4 and higher. The summation in
eq.(\ref{reference}) runs over the chirality index $p$ which is equal 1 for
left-movers and -1 for right-movers.

As we have mentioned above the colons denote the normal ordering with
respect to the noninteracting ground state $\left| 0 \right>$.
Specifically, the normal order is defined by the equations:
\begin{eqnarray}
\psi^\dagger_p (x) \psi^{\vphantom{\dagger}}_p (y) =
\ \colon\! \psi^\dagger_p (x) \psi^{\vphantom{\dagger}}_p (y)\colon
+ s_p (x-y), \label{normal1} \\
\psi^\dagger_p (x) \psi^{\vphantom{\dagger}}_p (y) 
\psi^\dagger_p (x') \psi^{\vphantom{\dagger}}_p (y') =
\ \colon\! \psi^\dagger_p (x) \psi^{\vphantom{\dagger}}_p (y) 
\psi^\dagger_p (x') \psi^{\vphantom{\dagger}}_p (y') \colon \label{normal2}\\
+ s_p (x-y) \colon\! \psi^\dagger_p (x') \psi^{\vphantom{\dagger}}_p (y')
\colon +
s_p (x'-y') \colon\! \psi^\dagger_p (x) \psi^{\vphantom{\dagger}}_p (y)
\colon \nonumber \\
+ s_p (x-y') \colon\! \psi^{\vphantom{\dagger}}_p (y) \psi^\dagger_p (x')
\colon +
s_p (y-x') \colon\! \psi^\dagger_p (x) \psi^{\vphantom{\dagger}}_p (y')
\colon \nonumber\\
+ s_p (x-y) s_p (x'-y') + s_p (x-y') s_p(y-x'),\nonumber \\
s_p(x) = \left< 0\right| \psi^\dagger_p (x) \psi^{\vphantom{\dagger}}_p (0)
\left| 0 \right>= \label{s}\\
\left< 0 \right| \psi^{\vphantom{\dagger}}_p (x) \psi^\dagger_p (0)
\left|0 \right>=
 \frac{p}{2\pi i\left(x-ip0\right)}. \nonumber 
\end{eqnarray}
Obviously, the Hamiltonian (\ref{reference}) is trivially soluble. The
spectrum of the fermionic excitations is
\begin{eqnarray}
\tilde \varepsilon_{pq} = p \tilde v_{\rm F} q + \tilde v_{\rm F}' q^2 +
{O}(q^3).
\end{eqnarray}
The contribution ${O}(q^3)$ comes from the terms of eq. (\ref{reference})
denoted by the ellipsis. These terms guarantee that $\tilde 
\varepsilon_{pq} = 0$ for $q=0$ only.

The density-density propagator for the model (\ref{reference}) 
\begin{eqnarray}
{\cal D}_{q\omega}^0 =  \frac{1} {4\pi \tilde v'_{\rm F} q}
\log \left\{ \frac{ \omega^2 + ( \tilde v_{\rm F} q -
\tilde v'_{\rm F} q^2)^2}{ \omega^2 + (\tilde v_{\rm F} q +
\tilde v'_{\rm F} q^2)^2 }\right\}\label{Df}
\end{eqnarray}
can be calculated without much effort.

Now we define the unitary transformation $U$:
\begin{eqnarray}
U={\rm e}^{\Omega},\label{U}\\
{\Omega} = -\frac{1}{2} \sum_{q \ne 0 } \sum_p p
\frac{\alpha_q}{n_q} \rho_{p-q} \rho_{-pq},\label{Omega}\\
n_q = \frac{Lq}{2\pi},\quad  \rho_{pq} = \int dx {\rm e}^{-iqx} \rho_p (x),
\end{eqnarray}
and transform with its help the operator $H_0$:
\begin{eqnarray}
H_{\rm ex} = U H_0 U^\dagger.
\end{eqnarray}
The subscript `ex' stands for `exact' since the Hamiltonian $H_{\rm ex}$ is
unitary equivalent to the quadratic Hamiltonian (\ref{reference}) which
implies exact solubility of $H_{\rm ex}$.

The coefficients $\alpha$'s from eq.(\ref{Omega}) uniquely specify the
operator $U$. We impose on them the following conditions:
\begin{eqnarray}
\alpha_q = \alpha_{-q},\\
\alpha_q \approx \alpha_q\big|_{q=0}{\rm \ for\ } |q| < \Lambda,\\
\alpha_q \rightarrow 0 {\rm \ for\ } |q| \gg \Lambda. \label{alpha}
\end{eqnarray} 
These conditions mean that $U$ when acting in the Fock space of Hamiltonian
(\ref{reference}) cannot create highly excited single-fermion states.
Loosely speaking, the action of $U$ is confined to the region $|q| <
\Lambda$ near the Fermi points. This guarantees that the
interactions of $H_{\rm ex}$ have proper ultraviolet cutoff of $\Lambda$.

In order to find $H_{\rm ex}$ explicitly we need the following commutation
rules:
\begin{eqnarray}
\left[ \rho_{pq}, \rho_{p'-q'} \right] = \delta_{pp'} \delta_{qq'} p n_q,
\label{rho_comm}\\
\left[\rho_{pq}^{\vphantom{\dagger}}, \psi^\dagger_{p'}(x) \right] =
\delta_{pp'} {\rm e}^{-iqx} \psi^\dagger_{p'}(x), \label{psi_comm}
\end{eqnarray}
which are proven in \cite{haldane,vD-S}. 

Next, let us study how operator $U$ acts on different
observables. With the help of eq.(\ref{rho_comm}) it is possible to show
that for $q \ne 0$:
\begin{eqnarray}
U \rho_{pq} U^\dagger = u(\alpha_q) \rho_{pq} + v(\alpha_q) \rho_{-pq},
\label{BT}\\
u(\alpha_q) = u_q = {\rm ch\ } \alpha_q,\ v(\alpha_q) = v_q =
{\rm sh\ }\alpha_q,\\
u_q^2 - v_q^2 = 1. \label{unity}
\end{eqnarray}
For zero modes we have:
\begin{eqnarray}
U N_p U^\dagger = N_p,\\
N_p = \rho_{pq} \big|_{q=0}.
\end{eqnarray}
Such transformation rule is a consequence of definition (\ref{Omega}), in
which $\Omega$ commutes with $N_p$.

It is convenient to define the following symbols:
\begin{eqnarray}
\rho^u_p (x) = \int dy \hat u(y) \rho_p (x-y) =
( \hat u * \rho_p) (x),\label{rho_u}\\
\rho^v_p (x) = \int dy \hat v(y) \rho_p (x-y) = ( \hat v * \rho_p) (x),
\label{rho_v}\\
\hat u(x) = L^{-1} \sum_q u_q {\rm e}^{iqx} = \int \frac{dq}{2\pi} u_q
{\rm e}^{iqx},\\
\hat v(x) = L^{-1} \sum_q v_q {\rm e}^{iqx} = \int \frac{dq}{2\pi} v_q {\rm
e}^{iqx}.
\end{eqnarray}
Here the asterisks stand for convolutions of two functions.

Due to eq.(\ref{alpha}) we have $u_q = 1$ and $v_q = 0$ for $|q| \gg \Lambda$.
Therefore, $\hat v(x)$ is well defined, whereas $\hat u(x)$ is singular:
\begin{eqnarray}
\hat u(x) = \delta(x) + \hat w(x),\\
\hat w(x) = L^{-1} \sum_q w_q {\rm e}^{iqx} = \int \frac{dq}{2\pi} w_q
{\rm e}^{iqx},\\
w_q = u_q - 1 \rightarrow 0 {\rm \ for\ }|q| \gg \Lambda,
\end{eqnarray}
where $\hat w(x)$ is well-defined.

Both $\hat v(x)$ and $\hat w(x)$ are even functions of $x$. They vary slowly
on the scale of $(1/\Lambda)$ and vanish for $|x| \gg (1/\Lambda)$.

We express the action of $U$ on $\rho_p (x)$ in the following manner:
\begin{eqnarray}
U \rho_p (x) U^\dagger = \rho_p^u (x) + \rho_{-p}^v (x) + \delta
\rho_{p0} \label{UrU} \\
= \rho_p (x) + \rho_p^w (x) + \rho_{-p}^v (x) + \delta\rho_{p0}, \nonumber \\
\rho^w_p = \hat w * \rho_p, \\
\delta \rho_{p0} =  \frac{1}{L}
\left[(1 - u_0) N_p - v_0 N_{-p}\right]  \\
= - \frac{1}{L} \left[ w_0 N_p + v_0 N_{-p} \right], \nonumber \\
u_0 = u_q|_{q=0},\ v_0 = v_q|_{q=0},\ w_0 = w_q|_{q=0}.
\end{eqnarray}
The obtained result for $U \rho U^\dagger$ can be used to find $U H_{\rm kin}
U^\dagger$. We prove in Appendix that:
\begin{eqnarray}
\frac{ip}{2} \left[ \colon\psi_p^\dagger(x)
(\nabla\psi_p^{\vphantom{\dagger}}(x))\colon  -
\colon(\nabla\psi_p^\dagger(x)) \psi_p^{\vphantom{\dagger}}(x)\colon\right]
\label{rhorho}\\
=\pi\lim_{y\rightarrow x} \left\{ \rho_p(x) \rho_p(y) - b_p(x-y)\right\},
\nonumber\\
b_p (x) = \left< 0 \right| \rho_p (x) \rho_p (0) \left| 0 \right> = s^2_p (x).
\end{eqnarray}
These formulas express the kinetic energy density in terms of the density
operators. Since we know how the latter are transformed under the action of
$U$ we determine $U H_{\rm kin} U^\dagger$ easily, as the derivation below
demonstrates.

First, introduce the notation:
\begin{eqnarray}
\colon \rho_p (x) \rho_p (y) \colon = \rho_p (x) \rho_p (y) - b_p (x-y),
\label{:rr:}\\
\colon \rho_p (x) \rho_p (y) \rho_p (z) \colon = 
\rho_p (x) \rho_p (y) \rho_p (z) - \label{:rrr:}\\
b_p (x-y) \rho_p (z) - b_p (x-z) \rho_p (y) - b_p (y-z) \rho_p (x). \nonumber 
\end{eqnarray}
In the bosonization framework the above equations define the normal ordering
of the bosonic operators. In this paper we do not introduce the
Tomonaga-Luttinger bosons. Thus, the symbols $\colon \rho \rho \colon$ and
$\colon \rho \rho \rho \colon$ must be understood as a shorthand notation
for expressions like eq.(\ref{:rr:}) and eq.(\ref{:rrr:}) without further
hidden field-theoretical meaning. The reason for using normal ordered products
is that such objects are well defined analytically at any values of $x,y$
and $z$ while unordered products have singularities when arguments
coincide (e.g, the unordered product $\rho_p(x) \rho_p (y)$ has
a second order pole at $x=y$).

The new notation allows us to reformulate eq.(\ref{rhorho}) 
in a more compact way:
\begin{eqnarray}
\frac{ip}{2} \left[ \colon \! \psi_p^\dagger
(\nabla\psi_p^{\vphantom{\dagger}})\colon  -
\colon \!(\nabla\psi_p^\dagger) \psi_p^{\vphantom{\dagger}}\colon\right]
= \pi \colon \! \rho_p^2  \colon. \label{rrHkin}
\end{eqnarray}
In the bosonization context this equality corresponds to the
bosonization of the kinetic energy \cite{haldane,vD-S}. 

Using eq.(\ref{rhorho}) and eq.(\ref{rrHkin}) we find:
\begin{eqnarray}
U H_{\rm kin} U^\dagger = \pi \tilde v_{\rm F} \sum_p
 \int dx U\colon \rho_p^2 \colon U^\dagger = \\
\pi \tilde v_{\rm F} \int dx \sum_p \lim_{y \rightarrow x}
U \left\{ \rho_p (x) \rho_p (y) - b_p(x-y) \right\}
U^\dagger. \nonumber 
\end{eqnarray}
The expression under the limit sign is equal to:
\begin{eqnarray}
U \left\{ \rho_p (x) \rho_p (y) -
b_p(x-y) \right\} U^\dagger = \\
\left( \rho_p^u (x) + \rho_{-p}^v (x) + \delta\rho_{p0}\right)
\left( \rho_p^u (y) + \rho_{-p}^v (y) + \delta\rho_{p0}\right) \nonumber\\
- b_p (x-y). \nonumber
\end{eqnarray}
In this equation we need to normal order $\rho$ operators according to
definition eq.(\ref{:rr:}):
\begin{eqnarray}
\rho_p^u (x) \rho_p^u (y) = \ \colon\rho_p^u (x) \rho_p^u (y) \colon +
( \hat u*b_p* \hat u)(x-y).\label{auxI}
\end{eqnarray}
It is easy to show with the help of eq.(\ref{unity}) that 
for any function $f(x)$
\begin{eqnarray}
(\hat u*f* \hat u)(x) = f(x) + ( \hat v*f* \hat v)(x). \label{f(x)}
\end{eqnarray}
Therefore, eq.(\ref{auxI}) may be written differently:
\begin{eqnarray}
\rho_p^u (x) \rho_p^u (y) =
\ \colon \rho_p^u (x) \rho_p^u (y) \colon +\\
b_p (x-y) + ( \hat v*b_{p}* \hat v) (x-y).\nonumber 
\end{eqnarray}
Finally:
\begin{eqnarray}
\rho^v_{-p} (x) \rho^v_{-p} (y) = \ \colon \rho_{-p}^v (x)
\rho_{-p}^v (y) \colon + ( \hat v*b_{-p}* \hat v)(x-y).
\end{eqnarray}
Collecting all terms together one finds:
\begin{eqnarray}
U H_{\rm kin} U^\dagger = \pi \tilde v_{\rm F} \int dx \sum_p
\colon (\rho_p^u +
\rho_{-p}^v + \delta \rho_{p0} )^2 \colon  \label{UHkinU}\\
+2\pi c \tilde v_{\rm F}\Lambda^2 L, \nonumber
\end{eqnarray}
where $c$ is the dimensionless non-universal $c$-number:
\begin{eqnarray} 
c = \Lambda^{-2}( \hat v*(b_{\rm L} + b_{\rm R})* \hat v)(0) =
\int \frac{dq}{(2\pi \Lambda)^2} |q| v_q^2. \label{c}
\end{eqnarray}
In eq.(\ref{UHkinU}) contributions proportional to $\delta \rho_{p0}$ or
$(\delta \rho_{p0})^2$ are ${O}(L^{-1})$. Consequently, they may be
neglected.  The additive constant (the term proportional to $c$) can be
neglected as well. With this in mind we determine:
\begin{eqnarray}
U H_{\rm kin} U^\dagger = \label{Hkin}\\
\pi \tilde v_{\rm F} \int dx \left[ \sum_p
\left\{\colon \!(\rho_p^u )^2 \colon +
\colon \!(\rho_p^v )^2 \colon \right\} +
2\rho_{\rm L}^u  \rho_{\rm R}^v +
2\rho_{\rm L}^v  \rho_{\rm R}^u \right]. \nonumber 
\end{eqnarray}
The expression in the curly brackets has to be transformed:
\begin{eqnarray}
\int dx \left\{\colon \!(\rho_p^u )^2 \colon +
\colon \!(\rho_p^v )^2 \colon \right\} = \int dx \left\{ \colon \rho_p^2
\colon + 2 \colon (\rho_p^v)^2 \colon \right\}.
\end{eqnarray}
The second term can be written as:
\begin{eqnarray}
2\int dx \ \colon (\rho^v_p)^2 \colon =\label{rv} \\
2\int dx dx' dx''  \hat v(x-x')  \hat v(x-x'') \ \colon \rho_p (x')
\rho_p (x'') \colon = \nonumber\\
2\int dx' d x'' ( \hat v* \hat v)(x'-x'') \ \colon \rho_p(x')
\rho_p(x'') \colon = \nonumber\\
2 c_0^v \int dx \ \colon \rho_p^2 \colon +  h_{\rm int}^{(4)}. \nonumber 
\end{eqnarray}
The term $ h_{\rm int}^{(4)}$ is a sum of irrelevant operators whose
scaling dimensions are 4 and higher. That is why we put such superscript at
$h$:
\begin{eqnarray}
h_{\rm int}^{(4)} = 2 \sum_{n=1}^{\infty} c_{n}^v \Lambda^{-2n}
\int dx \ \colon \rho_p (\nabla^{2n} \rho_p )\colon, \\
c_n^v = \frac{\Lambda^{2n}}{(2n)!}\int dx x^{2n} ( \hat v* \hat v)(x),\\
c_0^v = \int dx ( \hat v* \hat v)(x) = v_0^2.
\end{eqnarray}
We establish:
\begin{eqnarray}
\int dx \left\{\colon \!(\rho_p^u )^2 \colon +
\colon \!(\rho_p^v )^2 \colon \right\} = \int dx ( 1 + 2 c^v_0 )
\colon \rho_p^2 \colon +  h_{\rm int}^{(4)}  \\
= ( 1 + 2 c^v_0 ) \frac{ip}{\pi} \int dx \ \colon \psi^\dagger_p \nabla
\psi^{\vphantom{\dagger}}_p \colon +  h_{\rm int}^{(4)}. \nonumber 
\end{eqnarray}
Here we used eq.(\ref{rrHkin}) to transform $\colon \rho^2_p \colon$.

The $\colon \rho_{\rm L} \rho_{\rm R} \colon$ term of eq.(\ref{Hkin}) can be
written as:
\begin{eqnarray}
2\pi \tilde v_{\rm F}
\int dx ( \rho^u_{\rm L} \rho^v_{\rm R} + \rho^v_{\rm L} \rho^u_{\rm R} ) =
\\
\int dx dx' \hat g(x-x') \rho_{\rm L} (x) \rho_{\rm R} (x'), \nonumber \\
\hat g(x) = 4\pi \tilde v_{\rm F} (v*u)(x) \Leftrightarrow
g_q = 4\pi \tilde v_{\rm F} v_q u_q. \label{g}
\end{eqnarray}
Ultimately, we obtain the following expression for $U H_{\rm kin}
U^\dagger$:
\begin{eqnarray}
U H_{\rm kin} U^\dagger = i v_{\rm F}\int_{-L/2}^{L/2} dx
\left( \colon\psi^\dagger_{{\rm L}} \nabla
\psi^{\vphantom{\dagger}}_{{\rm L}}\colon - \colon\psi^\dagger_{{\rm R}}
\nabla\psi^{\vphantom{\dagger}}_{{\rm R}}\colon \right) \label{Hqp}\\
+ \int dx dx' \hat g(x-x') \rho_{\rm L}(x) \rho_{\rm R}(x') + \pi \tilde
v_{\rm F}  h_{\rm int}^{(4)}  \nonumber\\
= H_{\rm TL} + \pi \tilde v_{\rm F}  h_{\rm int}^{(4)},\nonumber \\
v_{\rm F} = \tilde v_{\rm F} (1 + 2 c_0^v)
= \tilde v_{\rm F} (1 + 2 v_0^2). 
\end{eqnarray}
It is clear from eq.(\ref{Hqp}) that up to the linear combination of
irrelevant operators $ h^{(4)}_{\rm int}$ the classical
Tomonaga-Luttinger model is unitary equivalent to the model of free
fermions with the Hamiltonian $H_{\rm kin}$. This was shown in
\cite{rozhkov}.

At first sight it appears that our ability to map the Tomonaga-Luttinger
Hamiltonian on the free fermionic quasiparticles is incompatible with the
known fact that there is no quasiparticle pole in the Tomonaga-Luttinger
single-electron propagator. The resolution of this paradox is simple
\cite{rozhkov}: the overlap of the physical electron state and the
quasiparticle state is zero. Thus, the single-quasiparticle propagator 
\begin{eqnarray}
{\cal G}_{\rm qp} (x,\tau) = 
- \left< 0 \right| {\cal T}_\tau \left\{
\psi_p^{\vphantom{\dagger}} (x,\tau)  \psi^\dagger_p (0,0) \right\} 
\left| 0 \right>, \\
\psi_p (x,\tau) = {\rm e}^{\tau H_{\rm kin}} \psi_p (x)
{\rm e}^{-\tau H_{\rm kin}} ,
\end{eqnarray}
does have a pole but such an object has no physical significance. Because
of the orthogonality catastrophe the physically important single-electron
propagator
\begin{eqnarray}
{\cal G}_{\rm TL} (x,\tau) = - \left< 0_{\rm TL} \right| {\cal T}_\tau
\left\{ \psi_p^{\vphantom{\dagger}} (x,\tau)  \psi^\dagger_p (0,0) \right\} 
\left| 0_{\rm TL} \right>, \\
\psi_p (x,\tau) = {\rm e}^{\tau H_{\rm TL}} 
\psi_p (x)  {\rm e}^{-\tau H_{\rm TL}} ,\\
\left| 0_{\rm TL} \right> = U \left| 0 \right> - H_{\rm TL}\ 
{\rm ground \ state}, \label{gs}
\end{eqnarray}
is not connected to the quasiparticle propagator in a Fermi liquid manner:
${\cal G}_{\rm TL} (q,\omega) \ne Z{\cal G}_{\rm qp} (q,\omega) +
{\cal G}_{\rm reg} (q,\omega)$. The Tomonaga-Luttinger single-electron
propagator was derived
with the help of representation (\ref{Hqp}) in Ref.\cite{rozhkov}.

It is interesting to establish the relation between the Tomonaga-Luttinger
Hamiltonian parameters $v_{\rm F}$, $g$ and the free fermion parameter
$\tilde v_{\rm F}$.
If we define $g_0 = g_q|_{q=0} = 4\pi \tilde v_{\rm F} u_0 v_0$ we can write:
\begin{eqnarray}
\left( \frac{g_0}{2\pi} \right)^2 = 4 \tilde v_{\rm F}^2 
v^2_0 ( 1 + v^2_0 ) = v_{\rm F}^2 - \tilde v_{\rm F}^2 \Rightarrow\\
\tilde v_{\rm F} = v_{\rm F} \sqrt{ 1 - \left( 
\frac{g_0}{2\pi v_{\rm F}}\right)^2}. \label{vF}
\end{eqnarray}
The latter formula is very well known in the bosonization theory. It gives
the dependence of the plasmon velocity on the bare Fermi velocity $v_{\rm
F}$ and the interaction constant $g_0$. The latter interpretation does not
contradict to eq.(\ref{vF}) for the plasmon mode in the fermion system with
the Hamiltonian $H_{\rm kin}$ propagates with the velocity $\tilde v_{\rm F}$.

Our next task is to find $U H_{\rm nl} U^\dagger$. We prove in Appendix
that:
\begin{eqnarray}
\colon (\nabla \psi^\dagger_p )
(\nabla \psi^{\vphantom{\dagger}}_p ) \colon - \frac{1}{6} \nabla^2
\rho_p = \frac{4\pi^2}{3} \colon \rho_p^3 \colon.\label{rho3}
\end{eqnarray}
This expression may be familiar to those who work with the bosonization: it
is the bosonic form of $H_{\rm nl}$ \cite{haldane,kopietz_book}.

As a consequence of eq.(\ref{rho3}) one derives:
\begin{eqnarray}
U H_{\rm nl} U^\dagger = \frac{4\pi^2 \tilde v_{\rm F}'}{3}
\sum_p \int dx U \colon \rho_p^3 \colon U^\dagger.
\end{eqnarray}
The expression under the integration sign is transformed as follows:
\begin{eqnarray}
U \ \colon \rho_p^3(x) \colon U^\dagger =\label{UrrrU}
\\
\lim_{x' \rightarrow x} U \left\{
\colon \rho_p^2 (x) \colon \rho_p (x') - 2 b_p (x-x') \rho_p (x) \right\}
U^\dagger. \nonumber 
\end{eqnarray}
Using our result for $U \colon \rho_p^2 \colon U^\dagger$ we derive:
\begin{eqnarray}
U \left\{ \colon \rho_p^2 \colon \rho_p' - 2 b_p (x-x') \rho_p 
\right\} U^\dagger = \label{rrruv}\\
\left[ \colon \{\rho^{u}_p + \rho^v_{-p} 
+ \delta \rho_{p0} \}^2 \colon + c \Lambda^2 \right] 
\left\{ (\rho^u_p )' + (\rho_{-p}^v )' + \delta \rho_{p0} \right\}
\nonumber\\
- 2 b_p (x-x') \left\{ \rho^u_p + \rho_{-p}^v + \delta \rho_{p0}
\right\}. \nonumber 
\end{eqnarray}
In this equality $\rho_p' = \rho_p^{\vphantom{\dagger}} (x')$,
$(\rho_p^{u,v})' = \rho_p^{u,v} (x')$.

The cubic term in eq.(\ref{rrruv}) has to be normal ordered:
\begin{eqnarray}
\colon\! \{ \rho^u_p + \rho_{-p}^v + \delta \rho_{p0} \}^2 \colon\!
\{(\rho^u_p)' + (\rho_{-p}^v )' + \delta \rho_{p0} \} = \label{:rr:r}\\
\colon \{ \rho^u_p  + \rho_{-p}^v  + \delta \rho_{p0} \}^2 
\{ ( \rho^u_p )' + (\rho_{-p}^v )' + \delta \rho_{p0} \} \colon
\nonumber \\
+ 2 [(\hat u * b_p * \hat u) (x-x') + (\hat v * b_{-p} * \hat v) (x-x')]
\times \nonumber\\
\{ \rho^u_p + \rho_{-p}^v + \delta \rho_{p0} \}. \nonumber 
\end{eqnarray}
Next one has to substitute eq.(\ref{:rr:r}) into eq.(\ref{rrruv}) and collect
similar terms. Using eq.(\ref{f(x)}) and definition eq.(\ref{c}) it is easy
to show now that:
\begin{eqnarray}
U \colon \rho_p^3 \colon U^\dagger = \label{U:r3:U}\\
\colon \{ \rho^u_p + \rho_{-p}^v + \delta \rho_{p0} \}^3 \colon 
+ 3 c \Lambda^2 \{ \rho^u_p + \rho_{-p}^v + \delta \rho_{p0} \}. \nonumber 
\end{eqnarray}
Therefore, up to terms ${O} (L^0)$ we find:
\begin{eqnarray}
U H_{\rm nl} U^\dagger = \frac{4 \pi^2 \tilde v_{\rm F}'}{3} \sum_p \int dx
\left( \colon \{ \rho^u_p + \rho_{-p}^v  \}^3 \colon + 3 c \Lambda^2
 \rho_p \right).
\end{eqnarray}
Now we need to expand the third order binomial:
\begin{eqnarray}
\sum_p \colon \{ \rho^u_p + \rho_{-p}^v  \}^3 \colon = \label{bino}
\sum_p \left[ \colon (\rho^u_p)^3 \colon + \colon (\rho^v_p)^3 \colon
\right.\\
\left. + 3\ \colon (\rho_p^u)^2 \colon \rho_{-p}^v 
+ 3\ \colon (\rho_p^v)^2 \colon \rho_{-p}^u \right], \nonumber 
\end{eqnarray}
and process the terms one by one. The first term:
\begin{eqnarray}
\colon (\rho^u_p)^3 \colon = \ \colon ( \rho_p^{\vphantom{w}}
+ \rho^w_p )^3 \colon =\\
\ \colon \! \rho_p^3 \colon + 3 \colon \! \rho_p^2 \rho_p^w \colon +
3 \colon \! \rho_p^{\vphantom{w}} (\rho_p^w)^2 \colon +
\colon \! (\rho_p^w)^3 \colon. \nonumber 
\end{eqnarray}
We use the same trick which was exploited while obtaining eq.(\ref{rv}):
\begin{eqnarray}
\int dx\ \colon (\rho^u_p)^3 \colon = \ ( 1 + w_0)^3 \int dx\
\colon \rho_p^3 \colon + \\
\int dx \left[ \left\{ 3 \colon \! \rho_p^2 \rho_p^w \colon - 3 w_0 \colon
\rho_p^3 \colon \right\} + \right. \nonumber \\
\left.
\left\{ 3 \colon\! \rho_p^{\vphantom{w}} (\rho_p^w)^2 \colon -
3 w_0^2 \colon \rho_p^3 \colon \right\}
+ \left\{ \colon (\rho_p^w)^3 \colon - w_0^3 \colon \rho_p^3 \colon
\right\} \right]. \nonumber 
\end{eqnarray}
As with eq.(\ref{rv}) it is possible to convince ourselves that the
expressions in curly brackets have higher scaling dimension than the
scaling dimension of $\colon\!\rho_p^3\colon$. Indeed:
\begin{eqnarray}
\colon \rho_p^2 \rho^w_p \colon - w_0 \colon \rho_p^3 \colon
= \sum_{n=1}^{\infty} c_n^w \Lambda^{-2n}
 \colon \rho_p^2 \nabla^{2n} \rho_p \colon,\\
c_n^w = \frac{\Lambda^{2n}}{(2n)!} \int dx x^{2n} w(x).
\end{eqnarray}
Dimension counting shows that the operators under the summation sign have
the scaling dimensions of 5 and higher. This is bigger than 3, the dimension
of $\colon \! \rho_p^3\colon$.

Acting in a similar fashion we establish for the second, third and fourth
terms of eq.(\ref{bino}) that:
\begin{eqnarray}
\int dx \colon\! (\rho^v_p)^3 \colon = v_0^3 \int dx \colon \!\rho_p^3\colon
+ \int dx \left[ \colon\! (\rho^v_p)^3 \colon - v_0^3 \colon \!\rho_p^3
\colon \right],\\
\int dx \left( \colon (\rho_p^u)^2 \colon \rho_{-p}^v 
+ \ \colon (\rho_p^v)^2 \colon \rho_{-p}^u \right) = \\
(u_0^2 v_0^{\vphantom{\dagger}} + v_0^2 u_0^{\vphantom{\dagger}})
\int dx \colon \rho_p^2 \colon \rho_{-p}^{\vphantom{\dagger}} + \nonumber \\
\int dx \left( \left[ \colon (\rho_p^u)^2 \colon 
\rho_{-p}^v - u_0^2 v_0^{\vphantom{\dagger}}
\colon \rho_p^2\colon \rho_{-p}^{\vphantom{\dagger}}\right] +
\right.\nonumber \\
\left. \left[ \colon (\rho_p^v)^2 \colon \rho_{-p}^u -
v_0^2 u_0^{\vphantom{\dagger}}
\colon \rho_p^2\colon \rho_{-p}^{\vphantom{\dagger}}\right] \right),
\nonumber
\end{eqnarray}
where the expressions in square brackets has the scaling dimension 5 or
higher. 

Combining the above results we have:
\begin{eqnarray}
U H_{\rm nl} U^\dagger = \frac{4 \pi^2 v_{\rm F,ex}'}{3} \int dx \left(
\colon \rho^3_{\rm L} \colon + \colon \rho^3_{\rm R} \colon \right) +\\
2 \pi g'_{\rm ex} \int dx \left( \colon \rho_{\rm L}^2 \colon
\rho_{\rm R}^{\vphantom{\dagger}} + \colon \rho_{\rm R}^2 \colon
\rho_{\rm L}^{\vphantom{\dagger}} \right) + 
\frac{4 \pi^2 \tilde v_{\rm F}'}{3}  h_{\rm int}^{(5)}, \nonumber \\
v_{\rm F,ex}' = (u_0^3 + v_0^3) \tilde v_{\rm F}',\label{vF'}\\
g'_{\rm ex} = 2 \pi \tilde v_{\rm F}'
(u_0^2 v_0^{\vphantom{\dagger}} + v_0^2 u_0^{\vphantom{\dagger}}).
\label{g'ex}
\end{eqnarray}
The superscript in the notation $ h_{\rm int}^{(5)}$ reminds
us of the fact that $ h_{\rm int}^{(5)}$ is a series of operators
whose lowest scaling dimension is 5.

Further, combining eq.(\ref{rrHkin}) and eq.(\ref{rho3}) we write:
\begin{widetext}
\begin{eqnarray}
U H_{\rm nl} U^\dagger = 
\sum_p \int dx \left\{ v_{\rm F,ex}' \colon\nabla\psi^\dagger_p
\nabla \psi_p^{\vphantom{\dagger}} \colon +
i g'_{\rm ex} p  \left( \colon\psi_p^\dagger
(\nabla\psi_p^{\vphantom{\dagger}})\colon  -
\colon(\nabla\psi_p^\dagger) \psi_p^{\vphantom{\dagger}}\colon\right)
\rho_{-p} \right\} + 
\frac{4 \pi^2 \tilde v_{\rm F}'}{3} \left( h_{\rm int}^{(5)} +
3 c \Lambda^2 (N_{\rm L} + N_{\rm R}) \right).
\end{eqnarray}
Therefore, using the above formula and eq.(\ref{Hqp}) one finds the following
expression for $H_{\rm ex} = U H_0 U^\dagger$:
\begin{eqnarray}
H_{\rm ex} = i v_{\rm F}\int dx \left( \colon\psi^\dagger_{{\rm L}} \nabla
\psi^{\vphantom{\dagger}}_{{\rm L}}\colon - \colon\psi^\dagger_{{\rm R}}
\nabla\psi^{\vphantom{\dagger}}_{{\rm R}}\colon \right) + \label{Hexact}
\int dx dx' \hat g(x-x') \rho_{\rm L}(x) \rho_{\rm R}(x') +\\
v_{\rm F,ex}'\sum_p \int dx \colon\nabla\psi^\dagger_p
\nabla \psi_p^{\vphantom{\dagger}} \colon +
i g'_{\rm ex} \sum_p p \int dx  \left( \colon\psi_p^\dagger
(\nabla\psi_p^{\vphantom{\dagger}})\colon  -
\colon(\nabla\psi_p^\dagger) \psi_p^{\vphantom{\dagger}}\colon\right)
\rho_{-p} + \mu (N_{\rm L} + N_{\rm R}) +
\pi \tilde v_{\rm F} h_{\rm int}^{(4)} +
\frac{4 \pi^2 \tilde v_{\rm F}'}{3} h_{\rm int}^{(5)}.\nonumber 
\end{eqnarray}
\end{widetext}
Since at $v_{\rm F}' \ne 0$ the particle-hole symmetry is broken the bare
value of the chemical potential is shifted: in the formula
above $\mu = 4\pi^2 c \tilde v_{\rm F}' \Lambda^2$ whereas $\mu = 0$ in
Hamiltonian (\ref{reference}).

It is important to realize that $v_{\rm F,ex}'$, $g'_{\rm ex}$ and $g$ in
eq.(\ref{Hexact}) are not independent. By
dividing eq.(\ref{vF'}) on eq.(\ref{g'ex}) we obtain:
\begin{eqnarray}
\frac{v_{\rm F,ex}'}{g'_{\rm ex}} = \frac{1}{2\pi} \left(
\frac{u_0^2 + v_0^2}{u_0 v_0} - 1 \right).
\end{eqnarray}
Since
\begin{eqnarray}
\frac{u_0^2 + v_0^2}{u_0 v_0} = \frac{4 \pi v_{\rm F}}{g_0},
\end{eqnarray}
we establish:
\begin{eqnarray} 
{v_{\rm F,ex}'}= \left(\frac{2 v_{\rm F}}{g_0} - \frac{1}{2\pi}\right)
{g'_{\rm ex}}.
\label{line}
\end{eqnarray}
The deduced results can be summarized in the from of a theorem.

{\bf Theorem 1. }{\it Consider the Tomonaga-Luttinger model with nonlinear 
dispersion and the irrelevant operators $h_{\rm int}^{(4,5)}$ whose scaling
dimensions are 4 and higher. Such model is unitary
equivalent to a system of free fermions with Hamiltonian (\ref{reference})
provided that parameters $v_{\rm F,ex}'$, $g'_{\rm ex}$, $v_{\rm F}$ and
$g_0$ satisfy eq.(\ref{line}).}

The theorem trivially implies that the Tomonaga-Luttinger model with
Hamiltonian (\ref{Hexact}) whose
parameters satisfy relation (\ref{line}) is exactly soluble.
This allows one to find the density-density propagator of eq.(\ref{Hexact}).
Since the ground state $\left| 0_{\rm TL} \right>$ of Hamiltonian
(\ref{Hexact}) is given by eq.(\ref{gs}) we derive:
\begin{eqnarray}
\left< 0_{\rm TL} \right| \{
\rho_{{\rm L} q} (\tau) + \rho_{{\rm R}q} (\tau)\} 
\{\rho_{{\rm L} -q} (0) + \rho_{{\rm R} -q} (0) \}\left| 0_{\rm TL}
\right> = \\
\left< 0 \right| 
U^\dagger \{\rho_{{\rm L} q} (\tau) + \rho_{{\rm R} q} (\tau) \} U 
U^\dagger \{ \rho_{{\rm L} -q} (0) + \rho_{{\rm R} -q} (0) \} U 
\left| 0 \right>. \nonumber 
\end{eqnarray}
Here, by the Green's function definition, $\rho_{pq} (\tau) =
\exp (\tau H_{\rm ex}) \rho_{pq} \exp (-\tau H_{\rm ex})$, that is
the evolution is controlled by $H_{\rm ex}$, eq.(\ref{Hexact}). The action
of $U^\dagger = U^{-1}$ on the density and evolution operators is:
\begin{eqnarray}
U^\dagger ( e^{\tau H_{\rm ex}} \rho_{pq} e^{-\tau H_{\rm ex}} )U =
e^{\tau H_0} ( u_q \rho_{pq} - v_q \rho_{-pq} ) e^{-\tau H_0} .
\end{eqnarray}
Note the minus sign in the above transformation law. It differs from
eq.(\ref{BT}) since here we apply $U^\dagger$ rather than $U$. Observe as
well that the evolution of the density operators on the right-hand side of
the above equation is set by the Hamiltonian (\ref{reference}). Thus:
\begin{eqnarray}
\left< 0_{\rm TL} \right| \{
\rho_{{\rm L} q} (\tau) + \rho_{{\rm R}q} (\tau)\} 
\{\rho_{{\rm L} -q} (0) + \rho_{{\rm R} -q} (0) \}\left| 0_{\rm TL}
\right> = \\
(u_q - v_q)^2 
 \left< 0 \right| \{\rho_{{\rm L} q} + \rho_{{\rm R} q} \}
e^{-\tau H_0} \{ \rho_{{\rm L} -q} +
\rho_{{\rm R} -q}  \}  \left| 0 \right>. \nonumber 
\end{eqnarray}
The latter expectation value is easy to find. It is just a product of two
free single-quasiparticle propagators corresponding to Hamiltonian 
(\ref{reference}). Therefore, Matsubara propagator for the
Tomonaga-Luttinger model (\ref{Hexact}) is equal to:
\begin{eqnarray}
{\cal D}_q^{\rm ex} (\tau) = 
(u_q - v_q)^2 {\cal D}_q^0 (\tau),\label{Dm}
\end{eqnarray}
where ${\cal D}^0$ is the propagator for the Hamiltonian $H_0$ specified
by eq.(\ref{Df}). At low $|q|$ this formula may be cast in a traditional for
the bosonization literature way:
\begin{eqnarray}
{\cal D}_{q\omega}^{\rm ex} = {\cal K} {\cal D}_{q\omega}^0,\\
{\cal K} = (u_0 - v_0)^2 = \sqrt{\frac{2\pi
v_{\rm F} - g_0}{2\pi v_{\rm F} + g_0}}.
\end{eqnarray}
Here ${\cal K}$ is the usual Tomonaga-Luttinger liquid parameter.

The retarded propagator $D_{q\omega}^{\rm ex}$ and the spectral density 
$B_{q\omega}^{\rm ex} = - 2 {\rm Im\/} D_{q\omega}^{\rm ex}$ are:
\begin{eqnarray}
D_{q\omega}^{\rm ex} = \frac{{\cal K}} {4\pi \tilde v'_{\rm F} q}
\log \left\{ \frac{ ( \tilde v_{\rm F} q -
\tilde v'_{\rm F} q^2)^2 - (\omega + i0)^2}{ (\tilde v_{\rm F} q +
\tilde v'_{\rm F} q^2)^2 - (\omega +i0)^2}\right\}, \label{Dret} \\
B_{q\omega}^{\rm ex} = \frac{{\cal K}} {2 \tilde v'_{\rm F} q} \left\{
\vartheta \left( \omega^2 - (\tilde v_{\rm F} q - \tilde v'_{\rm F} q^2)^2
\right) - \right. \label{B}\\
\left.
\vartheta \left( \omega^2 - (\tilde v_{\rm F} q + \tilde v'_{\rm F} q^2)^2
\right)
\right\} {\rm sgn\ } \omega. \nonumber 
\end{eqnarray}
Due to exact solubility the spectral density is zero everywhere outside the
interval $\tilde v_{\rm F} |q| - |\tilde  v_{\rm F}'|q^2 < |\omega| <
\tilde v_{\rm F} |q| + |\tilde v_{\rm F}'|q^2$.

\section{Non-technical overview of the derivations}
\label{intuitive}

The presentation of the previous Section was rather technical. Below we
will give a less formal account of the derivations performed so far. 

It is obvious that the major source of technical complications are the zero
modes and the irrelevant operators $h^{(4,5)}$. At the same time, they are
the least important parts of the derived Hamiltonian (\ref{Hexact}). Thus, 
it is useful to redo the calculations of the Section II paying no
attention to the zero modes and highly irrelevant operators. That way we
will be able to capture easily the most important features of the algebraic
manipulations without losing too much significant information.

Since the material presented in this Section is a refashioning of the
discussion given above the reader should be prepared for
some degree of redundancy.

Our technical goal is to start with the Hamiltonian of the free fermions with
nonlinear dispersion $H_0$, eq.(\ref{reference}), and transform
$H_0$ into $H_{\rm ex}$, eq.(\ref{Hexact}). We begin the execution of this
program by writing $H_0$ as:
\begin{eqnarray}
H_0 = \int dx \sum_p \pi \tilde v_{\rm F} \colon\! \rho_p^2 (x) \colon +
\frac{4\pi^2 \tilde v_{\rm F}'}{3} \colon\! \rho_p^3 (x) \colon.
\label{bos}
\end{eqnarray}
The validity of this equation can be checked with the help of
eq.(\ref{rrHkin}) and eq.(\ref{rho3}). We already pointed out in the
previous Section that the above equation may be thought of as the bosonic
form of $H_0$ \cite{haldane,kopietz_book}.

As in Sect.\ref{exact} we want to transform $H_0$ with the help of $U$
and find $U H_0 U^\dagger$. Operator $U$ acts as a Bogoliubov rotation on
the density operators, thus we may write (see eq.(\ref{UrU}),
eq.(\ref{UHkinU}) and eq.(\ref{U:r3:U})):
\begin{eqnarray}
U \rho_p (x) U^\dagger = u_0 \rho_p (x) + v_0 \rho_{-p} (x) + \ldots,
\label{UrUsimp}\\
U \colon\! \rho_p^2 (x) \colon U^\dagger =
\colon \! (u_0 \rho_p (x) + v_0 \rho_{-p} (x))^2 \colon + \ldots, 
\label{UrrUsimp}\\
U \colon\! \rho_p^3 (x) \colon U^\dagger =
\colon \! (u_0 \rho_p (x) + v_0 \rho_{-p} (x))^3 \colon + \ldots, 
\label{UrrrUsimp}
\end{eqnarray}
where the ellipses stand for {\it (a)} additive constants, {\it (b)} zero modes
terms {\it (c)} irrelevant operators with high scaling
dimensions and {\it (d)} operators which are reduced to zero modes upon
integration over $x$. 
As it was
explained above such terms do nothing but cloud our view. Thus, we will
neglect them.

Applying the rules eq.(\ref{UrUsimp})-(\ref{UrrrUsimp}) to eq.(\ref{bos}) the
following is derived:
\begin{eqnarray}
U H_0 U^\dagger = \int dx \sum_p \pi \tilde v_{\rm F} \colon\! (u_0 \rho_p
+ v_0 \rho_{-p})^2 \colon \qquad\\
+ \frac{4 \pi^2 \tilde v_{\rm F}'}{3} \colon\!
(u_0 \rho_p + v_0 \rho_{-p} )^3 \colon + \ldots  \nonumber\\
= \int dx \sum_p \pi \tilde v_{\rm F} (u_0^2 + v_0^2) \colon\! \rho_p^2
\colon + \frac{4 \pi^2 \tilde v_{\rm F}'}{3} (u_0^3 + v_0^3) \colon\!
\rho_p^3 \colon \nonumber\\
+ 2\pi \tilde v_{\rm F} u_0 v_0 \rho_p \rho_{-p} + 4\pi^2 \tilde v_{\rm F}'
(u_0^2 v_0^{\vphantom{2}} + v_0^2 u_0^{\vphantom{2}}) \colon\! \rho_p^2
\colon \rho_{-p}^{\vphantom{2}} + \ldots, \nonumber 
\end{eqnarray} 
where the ellipsis stand for the zero modes and highly irrelevant operators.
One notes that the term $\colon\! \rho_p^2 \colon$ is proportional to
$ip \colon\! \psi_p^\dagger \nabla \psi_p^{\vphantom{\dagger}} \colon$ and,
thus, the coefficient in front of $\colon\! \rho_p^2 \colon$ becomes
$v_{\rm F}$: $v_{\rm F} = \tilde v_{\rm F} (u_0^2 + v_0^2)$. Likewise, the
coefficient of $\rho_{\rm L} \rho_{\rm R}$ becomes $g_0$ (see eq.(\ref{g})),
the coefficient in front of $\colon\! \rho_p^2\colon
\rho_{-p}^{\vphantom{\dagger}}$ is $g'_{\rm
ex}$ (see eq.(\ref{g'ex})) and the coefficient of
$\colon\! \rho_p^3\colon$ is $v_{\rm F,ex}'$ (see eq.(\ref{vF'})).
Consequently, we obtain:
\begin{eqnarray}
H_{\rm ex} = U H_0 U^\dagger =\\
\int dx \sum_p i p v_{\rm F}
\colon\! \psi_p^\dagger \nabla \psi_p^{\vphantom{\dagger}} \colon +
v_{\rm F, ex}' \colon\! \nabla \psi_p^\dagger \nabla
\psi_p^{\vphantom{\dagger}} \colon 
+ \frac{1}{2} g_0 \rho_p \rho_{-p} \nonumber \\
+ i p g'_{\rm ex}  \left( \colon\psi_p^\dagger
(\nabla\psi_p^{\vphantom{\dagger}})\colon  -
\colon(\nabla\psi_p^\dagger) \psi_p^{\vphantom{\dagger}}\colon\right)
\rho_{-p} + \ldots. \nonumber 
\end{eqnarray}
%
The previous equation coincide with eq.(\ref{Hexact}) up to zero modes and
highly irrelevant operators. The coefficients $g'_{\rm ex}$ and $v_{\rm F}'$
satisfy eq.(\ref{line}). This concludes our nontechnical overview of the
exactly soluble Hamiltonian derivation. Transcending many technical details
discussed in Section II we discover that the method of calculating
$H_{\rm ex}$ is in fact fairly simple.

Another question we want to discuss here is the meaning of the exact
solubility line eq.(\ref{line}). Consider the generic Hamiltonian 
$H_{\rm G}$:
\begin{eqnarray}
H_{\rm G} = H_{\rm TL} + H_{\rm nl} + H_{\rm int}',
\end{eqnarray}
whose parameters do not satisfy any specific relation. It is possible to
find the transformation $V$ of the form eq.(\ref{U}) which diagonalizes the
first term of the above equation. For the detailed calculations
one should consult Ref.\cite{rozhkov} or our derivation of eq.(\ref{Hqp}). The heuristic explanation, however,
is easy. Consider the `bosonic' form of $H_{\rm TL}$:
\begin{eqnarray}
H_{\rm TL} = 
\frac{\pi v_{\rm F}}{L} \sum_{pq} \left(\rho_{pq} \rho_{p-q} +
\frac{g_q}{2\pi v_{\rm F}} \rho_{pq} \rho_{-p-q} \right) + \ldots
\end{eqnarray}
Application of $V$ produces the following:
\begin{eqnarray}
V H_{\rm TL} V^\dagger = \\
\frac{\pi v_{\rm F}}{L} \sum_{pq } \left[ u_q^2 + v_q^2 +
\frac{g_q}{\pi v_{\rm F}} u_q v_q \right] \rho_{pq} \rho_{p-q} +
\nonumber\\
\left[ \frac{g_q}{2\pi v_{\rm F}} (u_q^2 + v_q^2) + 2 u_q v_q \right]
\rho_{pq}\rho_{-p-q} + \ldots \nonumber 
\end{eqnarray}
By adjusting $u_q$ and $v_q$ we can kill the interaction term
$\rho_{p}\rho_{-p}$. For this to happens $u,v$ must satisfy the equation:
\begin{eqnarray}
\frac{g_q}{2\pi v_{\rm F}} (u_q^2 + v_q^2) + 2 u_q v_q = 0,
\end{eqnarray}
or, equivalently:
\begin{eqnarray}
{\rm th\ } 2\alpha_q = -\frac{g_q}{2\pi v_{\rm F}}. \label{alpha_def}
\end{eqnarray}
If this condition is met we have:
\begin{eqnarray} 
V H_{\rm TL} V^\dagger = 
i{\tilde v_{\rm F}} \int dx \left( \colon\psi^\dagger_{{\rm L}}
\nabla\psi^{\vphantom{\dagger}}_{{\rm L}}\colon - \colon\psi^\dagger_{{\rm R}}
\nabla\psi^{\vphantom{\dagger}}_{{\rm R}}\colon \right) + \ldots,
\label{Htldiag}
\end{eqnarray}
where the product $\rho_p \rho_p$ was expressed in terms of the fermionic
operators $\psi, \psi^\dagger$ with the help of eq.(\ref{rrHkin}).

There exists another transformation $V'$ of the general form (\ref{U}) but
with a different set of coefficients $\alpha'$ which diagonalizes the sum
$(H_{\rm nl} + H_{\rm int}')$:
\begin{eqnarray}
V' (H_{\rm nl} + H_{\rm int}') (V')^\dagger = \tilde v_{\rm F}'
\int dx \sum_p 
\colon 
       \nabla \psi_p^\dagger \nabla \psi_p^{\vphantom{\dagger}}
\colon 
+ \ldots
\end{eqnarray}
For this to take place the parameters $\alpha'$ must satisfy the relation:
\begin{eqnarray}
2\pi v_{\rm F}' + g' ( 1 + 2 {\rm cth\ } 2\alpha'_0 ) = 0,
\label{alpha'_def}
\end{eqnarray}
where $\alpha_0'$ equals to $\alpha_q'$ at $q=0$. The derivations of the two
above equations is similar to the proof of equations (\ref{alpha_def}) and
(\ref{Htldiag}). The idea of these derivations is to write 
$V'(H_{\rm nl} + H_{\rm int}')(V')^\dagger$ in terms of $\rho$'s:
\begin{eqnarray}
V'(H_{\rm nl} + H_{\rm int}')(V')^\dagger = V' \Big[
\int dx \frac{4 \pi^2 v_{\rm F}'}{3}
(\colon\! \rho_{\rm L}^3 \colon + \colon\! \rho_{\rm R}^3 \colon ) \quad\\
+ 2\pi g' (\colon\! \rho_{\rm L}^2 \colon \rho_{\rm R} +
\colon\! \rho_{\rm R}^2 \colon \rho_{\rm L} ) + \ldots \Big] (V')^\dagger,
\nonumber 
\end{eqnarray} 
and adjust $\alpha_0'$ of $V'$ in such a way as to remove $:\rho_p^2:
\rho_{-p}^{\vphantom{\dagger}}$ terms. This requirement lead us to
eq.(\ref{alpha'_def}). Once the latter terms are removed what remains
is proportional to $:\rho^3_{\rm L}: + :\rho^3_{\rm R}:$. This expression is
proportional to $H_{\rm nl}$ and, therefore, is diagonal.
 
If the values of $v_{\rm F}$, $v_{\rm F}'$, $g$ and $g'$ are arbitrary then
$V \ne V'$. Consequently, the whole Hamiltonian $H_{\rm G}$ cannot be
diagonalized with the help of the canonical transformation of the form
eq.(\ref{U}). However, if we demand that $V = V'$ or, equivalently,
$\alpha = \alpha'$ such diagonalization becomes possible and the system
becomes soluble. One can check that the exact solubility condition
(\ref{line}) may be derived from the equation $\alpha_0 = \alpha'_0$.
Thus, one can say that the exact
solubility condition guarantees the simultaneous diagonalization of the whole
Hamiltonian.

\section{Generic Tomonaga-Luttinger model with nonlinear dispersion}
\label{generic}

In this Section we will discuss how the above findings can be applied
to a generic model of the one-dimensional spinless fermions.

The most general Hamiltonian is:
\begin{eqnarray}
H_{\rm G} = H_{\rm TL} + H_{\rm nl} + H_{\rm int}' + H^{(4)}, \label{HG}
\end{eqnarray}
where $H^{(4)}$ stands for operators whose scaling dimensions are 4 and
higher. Here no special relation between $v_{\rm F}'$ and $g'$ is assumed.
Neither do we suppose that $H^{(4)}$ is equal to the linear combination
of $ h^{(4)}$ and $ h^{(5)}$ from eq. (\ref{Hexact}). To avoid clutter we do
not show the chemical potential term explicitly for it can be accounted
trivially.

It is convenient to write Hamiltonian $H_{\rm G}$ in the form:
\begin{eqnarray}
H_{\rm G} = H_{\rm ex} + \delta H_{\rm nl} + \delta H^{(4)}, \label{HG2}\\
\delta H_{\rm nl} = 
\delta v_{\rm F}' 
\int dx \sum_p 
\colon\! 
        \nabla\psi^\dagger_p \nabla \psi_p^{\vphantom{\dagger}} 
\colon,\\
\delta v_{\rm F}' = v_{\rm F}' - v_{\rm F,ex}'.
\end{eqnarray}
The term $\delta H^{(4)}$ contains those operators whose scaling dimension
are 4 and higher and which are not absorbed into $H_{\rm ex}$. The
constant $\delta v_{\rm F}'$ and $\delta H^{(4)}$ measure the deviation of
the Hamiltonian $H_{\rm G}$ from the exact solubility line. We want to apply
the transformation $V$ of the form eq.(\ref{U}) to $H_{\rm G}$
such that the transformed Hamiltonian contains no marginal interaction
operator $H_{\rm int}$. The remaining irrelevant operators can be treated
perturbatively.  

We already know how the first two terms of eq.(\ref{HG2})
are transformed under the action of $V$. The unknown entity is
$V \delta H^{(4)} V^\dagger$. To discuss the action of $V$ on
$\delta H^{(4)}$ it is useful to introduce the notation:
\begin{eqnarray}
h_{\rm kin} = i \int dx
\left( \colon\psi^\dagger_{{\rm L}}
\nabla\psi^{\vphantom{\dagger}}_{{\rm L}}\colon - \colon\psi^\dagger_{{\rm
R}}
\nabla\psi^{\vphantom{\dagger}}_{{\rm R}}\colon \right) ,\\
h_{\rm int} = \int dx \rho_{\rm L} \rho_{\rm R},\\
h_{\rm int}' = \int dx \sum_p 
i p  \rho_{-p} \left( \colon \psi^\dagger_p
(\nabla \psi^{\vphantom{\dagger}}_p) \colon -
\colon (\nabla \psi^\dagger_p) \psi^{\vphantom{\dagger}}_p
\colon \right),\\
h_{\rm nl} = \int dx \sum_p \colon\nabla\psi^\dagger_p
\nabla \psi_p^{\vphantom{\dagger}} \colon. 
\end{eqnarray} 
These operators exhaust the list of operators whose scaling dimension is
lower than 4 and we can write the most general expression:
\begin{eqnarray}
V \delta H^{(4)} V^\dagger = \\
 \beta_{\rm kin} h_{\rm kin} + \beta_{\rm int}
 h_{\rm int} + \beta_{\rm nl} h_{\rm nl} + \beta_{\rm int}' h_{\rm int}' 
+ \delta \bar H^{(4)}. \nonumber
\end{eqnarray}
The coefficients $\beta$'s are functions of $\alpha$'s which specify
$V$. 

Now we must consider two cases: {\it (i)} when $\beta_{\rm int} = 0$ and
{\it (ii)} when $\beta_{\rm int} \ne 0$. 

Let us start with {\it (i)}. The transformation $V$ with
\begin{eqnarray}
\alpha_0 = -\frac{1}{2} {\rm th}^{-1} \left( \frac{g_0}{2\pi v_{\rm F}}
\right) \label{a0}
\end{eqnarray}
acts on $H_{\rm G}$ as follows:
\begin{eqnarray}
V H_{\rm G} V^\dagger = H_0 + (\delta \tilde v_{\rm F}' h_{\rm nl} + 
\delta \tilde g' h_{\rm int}' ) + V \delta H^{(4)} V^\dagger \label{VHGVI}\\
= \left[ H_0 + \beta_{\rm kin} h_{\rm kin} +
(\delta \tilde v_{\rm F}' + \beta_{\rm nl}) h_{\rm nl} \right] 
\nonumber\\
+ (\delta \tilde g' + \beta_{\rm int}') h_{\rm int}' 
+ \delta \bar H^{(4)}, \nonumber \\
\delta \tilde v_{\rm F}' = (u_0^3 - v_0^3) \delta v_{\rm F}' ,\\
\delta \tilde g' = 2\pi (v^2_0 u_0 - u_0^2 v_0) \delta v_{\rm F}'.
\end{eqnarray}
In the first line of eq.(\ref{VHGVI}) the expression in the round brackets
corresponds to $V \delta v_{\rm F}' h_{\rm nl} V^\dagger$.

Eq.(\ref{VHGVI}) can be reformulated in a more compact way:
\begin{eqnarray}
V H_{\rm G} V^\dagger = \bar H_0 + \delta H_{\rm irr}, \label{VHGV}\\
\bar H_0 = H_0 + \beta_{\rm kin} h_{\rm kin} +
(\delta \tilde v_{\rm F}' + \beta_{\rm nl}) h_{\rm nl},\\
\delta H_{\rm irr} = (\delta \tilde g' + \beta_{\rm int}') h_{\rm int}' 
+ \delta \bar H^{(4)}.
\end{eqnarray}
Hamiltonian $\bar H_0$ is quadratic in the fermionic field operators. It
is similar to $H_0$ but has renormalized parameter values. Likewise,
$\delta \bar H^{(4)}$ is the renormalized version of $\delta H^{(4)}$.

We see that deviation from the exact solubility line eq.(\ref{line}) in
$\delta v_{\rm F}'$ has two effects: {\it (a)} renormalization of
$\tilde v_{\rm F}$ and $\tilde v_{\rm F}'$ and {\it (b)} introduction of
the interactions between the quasiparticles (operator $\delta H_{\rm irr}$).
Fortunately, these interactions are irrelevant in the renormalization group
sense. Thus, their influence could be accounted perturbatively. For that
reason let us introduce the notation: 
\begin{eqnarray}
H_{\rm qp}=V H_{\rm G} V^\dagger,
\end{eqnarray} 
which suggests that the transformed operator $V H_{\rm G} V^\dagger$
represents the Hamiltonian of the weakly interacting quasiparticles.

Next we study the case {\it (ii)}. This situation is not much more
difficult than {\it (i)}. From the field-theoretical point of view
$\beta_{\rm int} \ne 0$ means that irrelevant operators contribute to the
renormalization of the marginal operator coupling constant $g$. The correct
way to address the problem is to fix $\alpha_0$ by a generalized version of
eq.(\ref{a0}). The value of $\alpha_0$ must be determined by the
requirement that the quasiparticle Hamiltonian 
\begin{eqnarray}
H_{\rm qp} =
\gamma_{\rm kin} h_{\rm kin} + \gamma_{\rm int}
 h_{\rm int} + \gamma_{\rm nl} h_{\rm nl} + \gamma_{\rm int}' h_{\rm int}'
\label{VHGV2}\\
+ \delta \bar H^{(4)} \nonumber 
\end{eqnarray}
($\gamma$'s are functions of $\alpha_0$) contains no $h_{\rm
int}$:
\begin{eqnarray}
\gamma_{\rm int} (\alpha_0) = 0. \label{a0_gen}
\end{eqnarray}
It is easy to demonstrate that eq.(\ref{a0_gen}) is equivalent to
eq.(\ref{a0}) if $\beta_{\rm int} = 0$.

Once $\alpha_0$ is determined according to eq.(\ref{a0_gen}) the expression
(\ref{VHGV2}) attains the form of eq.(\ref{VHGV}). Thus, we prove

{\bf Theorem 2. }{\it A generic Tomonaga-Luttinger model with nonlinear
dispersion is unitary equivalent to a system of the free quasiparticles with
irrelevant interactions.}

\section{Calculations of observables}
\label{observe}

What does Theorem 2 imply for evaluation of the physical properties of
$H_{\rm G}$?  The irrelevance of corrections $\delta H_{\rm irr}$ means that
the ground state properties of Hamiltonian $H_{\rm G}$ can be
approximated by the ground state properties of $\bar H_0$. Should we be
interested in corrections those could be found perturbatively.

Further, for an observable ${\cal O}(x)$ the correlation function
\begin{eqnarray}
{\cal G}_{\cal O} =
- \left< 0_{\rm G} \right| {\cal T}_\tau \left\{ {\cal O}(x,\tau) 
{\cal O}^\dagger (0,0) \right\} \left| 0_{\rm G} \right>,\\
{\cal O} (x,\tau) = {\rm e}^{\tau H_{\rm G}} {\cal O}(x) 
{\rm e}^{- \tau H_{\rm G}},\\
\left| 0_{\rm G} \right> - H_{\rm G} {\rm \ ground \ state}, 
\end{eqnarray}
may be expressed as:
\begin{eqnarray}
{\cal G}_{\cal O} = 
- \left< 0_{\rm qp} \right| {\cal T}_\tau
\left\{ \tilde {\cal O}(x,\tau) \tilde {\cal O}^\dagger (0,0) \right\}
\left|  0_{\rm qp} \right>,\label{propa} \\
\tilde {\cal O} (x,\tau) = {\rm e}^{\tau  H_{\rm qp} }
\left( V {\cal O}(x) V^\dagger \right) {\rm e}^{- \tau  H_{\rm qp} }, \\
\left| 0_{\rm qp} \right> - {\rm \ ground \ state\ of\ }
 H_{\rm qp}  . 
\end{eqnarray}
If the observable ${\cal O}$ is such that $V {\cal O} V^\dagger$ has a very
complicated form the practical evaluation of eq.(\ref{propa}) may be
difficult to carry out. An example of such situation is ${\cal O} \equiv
\psi_p$. At present the author does not know how the single-fermion
propagator for the case of nonzero $v_{\rm F}'$ can be calculated.

On the other hand, if the object $V {\cal O} V^\dagger$ has some simple
representation in terms of the quasiparticle degrees of freedom one can
evaluate ${\cal G}_{\cal O}$ with the help of the usual perturbative expansion
in orders of $\delta H_{\rm irr}$. For example, consider calculation of the
density-density propagator: ${\cal O} \equiv (\rho_{\rm L} + \rho_{\rm R})$.
In that instance
$V {\cal O} V^\dagger$ is a linear combination of the density operators.
The zeroth order expression for such propagator is:
\begin{eqnarray}
{\cal D}_{q\omega} = {\cal D}_{q\omega}^{\rm ex} + {O} ((\delta H_{\rm
irr})^2).
\end{eqnarray} 
The corrections in orders of $\delta H_{\rm irr}$ could be determined with the
help of field-theoretical techniques.

Note, that the actual calculation of ${O} ((\delta H_{\rm irr})^2)$
corrections may be quite nontrivial. Indeed, numerical evidence
\cite{pereira} and heuristic analytical arguments \cite{pust} indicates that
$D_{q\omega}$ of the generic model diverges stronger than $D_{q\omega}^{\rm
ex}$. This suggests that the perturbation theory for $D_{q\omega}$
requires a reliable resummation technique. 

However, we are not always
interested in the propagator itself. In fact, there are situations when a few
low-order diagrams contributing to $D_{q\omega}$ would suffice. For example,
consider some quantity $r$ proportional to an integral over $D_{q\omega}$.
[An important example of such quantity is the Coulomb drag resistivity
\cite{glazman}: $r \propto \int dq d\omega f(q,\omega) ({\rm Im\ }
D_{q\omega})^2$, where $f$ is some temperature-dependent kernel.] Unlike the
propagator itself, $r$ remains finite. This is so because the divergences of
$D_{q\omega}$ due to ${O} ((\delta H_{\rm irr})^2)$ corrections are very
weak. Thus, the integration of $D_{q\omega}$ can be successfully performed
with the result:
\begin{eqnarray}
r = r^{\rm ex} + \delta r,
\end{eqnarray}
where quasiparticle interaction correction $|\delta r| < \infty$ may be
non-analytic.

Regarding the quasiparticle interaction corrections due to
$\delta H_{\rm irr}$ we would like to make
a remark. Operator $(\delta \tilde g' + \beta_{\rm int}') h'_{\rm int}$ is
the least irrelevant part of $\delta H_{\rm irr}$. Consequently, we expect
that at small momenta the corrections due to 
$(\delta \tilde g' + \beta_{\rm int}') h'_{\rm int}$ are stronger than those
due to $\delta \bar H^{(4)}$. Therefore, the most important contribution to
the low momenta corrections are $\propto (\delta \tilde g' +
\beta_{\rm int}')^2$. Thus, there is a hypersurface in 
the generic one-dimensional model parameter space where the low momenta
corrections drastically weakens. This hypersurface is fixed by the equation 
\begin{eqnarray}
\delta \tilde g' + \beta_{\rm int}' = 0. \label{line2}
\end{eqnarray}
We can interpret this as a manifestation of the exact solubility line
eq.(\ref{line}). 

As a practical application one can calculate the hypersurface (\ref{line2})
for the spinless Hubbard-like model
\begin{eqnarray}
H_{\rm hub} = \sum_{i,j} t(i-j) c^\dagger_i c^{\vphantom{\dagger}}_j +
\mu \sum_i n_i + u \sum_i n_i n_{i+1} \label{hubbard}
\end{eqnarray}
in the limit of small interaction $u \ll t$. Using the decomposition:
\begin{eqnarray}
c_i = \psi_{\rm L} (x_i) e^{-ik_{\rm F} x_i} +
\psi_{\rm R} (x_i) e^{ik_{\rm F} x_i},
\end{eqnarray}
one can calculate $v_{\rm F}$, $v_{\rm F}'$, $g$ and $g'$ in terms of
$t(i)$ and $u$. As soon as this task is done the equations for those
quantities must be substituted into eq.(\ref{line}) and the limit $u
\rightarrow 0$ is taken. This gives:
\begin{eqnarray}
v_{\rm F,ex}' = a v_{\rm F}  {\rm ctg\ } (k_{\rm F} a), \label{line3}
\end{eqnarray}
where $a$ is the lattice constant and the values of $v_{\rm F}$ and $v_{\rm
F, ex}'$ are functionals of $t(i)$. 

What are the phenomenological manifestations of eq.(\ref{line3})? As
explained above, on the surface fixed by the latter equation and in the
limit $u \ll t$ the
scattering of the quasiparticles is drastically reduced. As a consequence,
the propagator $D_{q\omega}$ calculated for the Hamiltonian satisfying
eq.(\ref{line3}) differs qualitatively from the propagator of the generic
Hamiltonian eq.(\ref{hubbard}). For example, the divergence of the
propagator due to quasiparticle scattering might be weaker. Such prediction
may be tested numerically.

\section{Comparison with other work}
\label{compare}

The issue of nonzero $v_{\rm F}'$ has a long history. Traditionally, it has
been approached within the framework of the bosonization
\cite{haldane,kopietz_book,sam}. The disadvantage of the bosonization is
that the perturbative expansion in orders of $v_{\rm F}'$ is highly
singular. 

To circumvent this latter difficulty the authors of Ref.\cite{piroo}
used the random phase approximation to find ${D}_{q\omega}$. Their
spectral function differs qualitatively from eq.(\ref{B}). We believe that
such a result is an artefact of the random phase approximation. 

Alternatively, in several recent papers \cite{glazman,teber}
the use of the `fermionic' methods was advocated. The idea behind them is
that away from the mass surface the perturbation theory in $g$ is
permissible without bosonization. That is, the authors chose to work with
the original fermions but at a price: the perturbation theory results cannot
be applied straightforwardly near the mass surface.

The proposed method was used to calculate the high-frequency tail of the
propagator. The latter technique is complimentary to the exact solution: it
is easily implemented at high frequency but it is problematic at the mass
surface where the exact solution works.

In Ref. \cite{pust} the heuristic argument
was put forward suggesting that the spectral function $B$ of $H_{\rm G}$
diverges at some line of $(q,\omega)$ plane. This finding is corroborated
qualitatively by the numerical work \cite{pereira}. How does this compare
against calculations presented here?

Obviously, our spectral function $B^{\rm ex}$ of the exactly soluble model
is finite. This does not contradict to Ref.\cite{pust,pereira}: the
propagator of the generic model $H_{\rm G}$ is affected by the quasiparticle
interaction which induces the extra divergence. Thus, the propagator
calculation requires an adequate resummation technique, as we pointed out in
the previous Section. We may hypothesize that the argumentation of Ref.
\cite{pust} may be directly applied to the quasiparticle Hamiltonian
$H_{\rm qp}$. Such `marriage' of the two approaches possesses a clear
advantage: both marginal and irrelevant interactions are accounted for to all
orders. Further speculations are to be postponed until thorough investigation.

To conclude, we have shown that the Tomonaga-Luttinger model with specific
parameter values is equivalent to the free fermions. At this point the
Tomonaga-Luttinger Hamiltonian is exactly soluble. Away from the exact
solubility the model may be mapped on a system of the fermions with irrelevant
interactions. Although, under such circumstances the model is not soluble
exactly, yet, it can be investigated with the help of the perturbation theory.

\section{Acknowledgements}

The author is grateful for the support provided by the Dynasty Foundation,
by RFBR through grant 03-02-16626 and by Russian federal
program ``Leading scientific schools" NSh-1694.2003.2.

\section{Appendix}

In this Appendix we prove certain relations required by the presentation in
the main text.

First, we derive eq.(\ref{rhorho}). Our initial step is to prove:
\begin{eqnarray}
\colon \! \rho_p (x) \rho_p (y) \colon = \label{proof_:rr:}\ 
\colon \! \psi_p^\dagger(x) \psi_p^{\vphantom{\dagger}}(x)
\psi_p^\dagger(y) \psi_p^{\vphantom{\dagger}}(y)\colon +\label{rr}
\nonumber \\
s_p(x-y) \colon \! \psi_p^{\vphantom{\dagger}}(x) \psi_p^\dagger(y)\colon +
s_p(x-y) \colon \!\psi_p^{\dagger}(x) \psi_p^{\vphantom{\dagger}}(y)\colon .
\end{eqnarray}
The proof goes as follows:
\begin{eqnarray}
\rho_p (x) \rho_p (y) = \
\colon \psi_p^\dagger(x) \psi_p^{\vphantom{\dagger}}(x) \colon
\colon \! \psi_p^\dagger(y) \psi_p^{\vphantom{\dagger}}(y) \colon = \\
s_p^2 (x-y) + \colon \! \psi_p^\dagger(x) \psi_p^{\vphantom{\dagger}}(x)
\psi_p^\dagger(y) \psi_p^{\vphantom{\dagger}}(y)\colon +  \nonumber\\
s_p(x-y) \colon \! \psi_p^{\vphantom{\dagger}}(x) \psi_p^\dagger(y)\colon +
s_p(x-y) \colon \!\psi_p^{\dagger}(x) \psi_p^{\vphantom{\dagger}}(y)\colon,
\nonumber 
\end{eqnarray}
where we used eq. (\ref{normal2}). Taking
into consideration eq.(\ref{:rr:}) and $b_p = s_p^2$ one obtains the
desired formula eq.(\ref{proof_:rr:}).

Once eq.(\ref{proof_:rr:}) is established eq.(\ref{rhorho}) follows. We
send $y \rightarrow x$ and note that 
$\colon \! \psi_p^\dagger(x) \psi_p^{\vphantom{\dagger}}(x)
\psi_p^\dagger(y) \psi_p^{\vphantom{\dagger}}(y)\colon = 0$ if $x=y$.
Indeed:
\begin{eqnarray}
&&\colon \! \psi_p^\dagger(x) \psi_p^{\vphantom{\dagger}}(x)
\psi_p^\dagger (y) \psi_p^{\vphantom{\dagger}}(y)\colon = \label{anticomm}\\
&&\qquad - \colon \! \psi_p^\dagger(y) \psi_p^{\vphantom{\dagger}}(x)
\psi_p^\dagger(x) \psi_p^{\vphantom{\dagger}}(y)\colon . \nonumber 
\end{eqnarray}
Since the quartic term vanishes we need to evaluate the quadratic terms of
eq.(\ref{proof_:rr:}) only. To calculate them at $x=y$ one can utilize
the L'Hospital's rule:
\begin{eqnarray}
\lim_{y \rightarrow x}
\left[ s_p(x-y)\left\{ \colon \! \psi_p^{\vphantom{\dagger}}(x) 
\psi_p^\dagger(y)\colon +
\colon \!\psi_p^{\dagger}(x) \psi_p^{\vphantom{\dagger}}(y)\colon \right\}
\right] = \\
\frac{ip}{2\pi}\left\{
\colon \! \psi_p^{\vphantom{\dagger}}(x) \nabla \psi_p^\dagger(x)\colon +
\colon \!\psi_p^{\dagger}(x) \nabla
\psi_p^{\vphantom{\dagger}}(x)\colon \right\}, \nonumber 
\end{eqnarray}
where we took into account that 
\begin{eqnarray}
\frac{\partial}{\partial y}\left( \frac{1}{s_p (x-y)} \right) =
\frac{2\pi}{ip}.
\end{eqnarray}
It is obvious now that eq.(\ref{rhorho}) is valid.

Next, we want to prove eq.(\ref{rho3}). The procedure is similar to the one
we have used just now. It is possible to show:
\begin{widetext}
\begin{eqnarray}
\colon \rho_p (x) \rho_p (y) \rho_p (z) \colon = 
\colon 
\psi_p^\dagger (x) \psi_p^{\vphantom{\dagger}} (x)
\psi_p^\dagger (y) \psi_p^{\vphantom{\dagger}} (y)
\psi_p^\dagger (z) \psi_p^{\vphantom{\dagger}} (z)\colon 
\label{proof_:rrr:} \\
+ s_p (x-y) \left\{\colon \psi^{\dagger}_p(x) \psi^{\vphantom{\dagger}}_p(y)
\psi^{\dagger}_p(z) \psi^{\vphantom{\dagger}}_p(z) \colon - 
\colon\psi^{\dagger}_p(y) \psi^{\vphantom{\dagger}}_p(x)
\psi^{\dagger}_p(z) \psi^{\vphantom{\dagger}}_p(z) \colon \right\} 
\nonumber \\
+ s_p (x-z) \left\{ \colon \psi^{\dagger}_p(x) \psi^{\vphantom{\dagger}}_p(z)
\psi^{\dagger}_p(y) \psi^{\vphantom{\dagger}}_p(y) \colon - 
\colon\psi^{\dagger}_p(z) \psi^{\vphantom{\dagger}}_p(x)
\psi^{\dagger}_p(y) \psi^{\vphantom{\dagger}}_p(y) \colon \right\}
\nonumber \\
+ s_p (y-z) \left\{\colon \psi^{\dagger}_p(y) \psi^{\vphantom{\dagger}}_p(z)
\psi^{\dagger}_p(x) \psi^{\vphantom{\dagger}}_p(x) \colon - 
\colon\psi^{\dagger}_p(z) \psi^{\vphantom{\dagger}}_p(y)
\psi^{\dagger}_p(x) \psi^{\vphantom{\dagger}}_p(x) \colon \right\}
\nonumber\\
+ s_p( y-z) s_p (x-z) \left\{
\colon \psi_p^{\vphantom{\dagger}} (x) \psi_p^{\dagger}(y) \colon -
\colon \psi_p^\dagger (x) \psi_p^{\vphantom{\dagger}}(y) \colon \right\} +
s_p( x-y) s_p (x-z) \left\{
\colon \psi_p^{\vphantom{\dagger}} (y) \psi_p^\dagger(z) \colon -
\colon \psi_p^\dagger (y) \psi_p^{\vphantom{\dagger}}(z) \colon \right\}
\nonumber \\
- s_p( x-y) s_p (y-z) \left\{
\colon \psi_p^{\vphantom{\dagger}} (x) \psi_p^\dagger(z) \colon -
\colon \psi_p^\dagger (x) \psi_p^{\vphantom{\dagger}} (z) \colon \right\}.
\nonumber 
\end{eqnarray}
\end{widetext}
Like eq.(\ref{proof_:rr:}), this equality was deduced by writing $\rho \rho
\rho$ as a product of six fermionic fields which is then expressed as a sum
of the normal ordered expressions. If we put in the above equation $x=y=z$
only terms quadratic in the fermionic fields would survive. This is a
consequence of the fermionic field anticommutativity (see
eq.(\ref{anticomm})).  Since we are interested in $x=y=z$ limit we have to
handle the quadratic part of eq.(\ref{proof_:rrr:}) only.

Let us define an operator-valued function:
\begin{eqnarray}
G(x,y,z) = \qquad \label{G}\\
s_p( y-z) s_p (x-z) \left\{
\colon \psi_p^{\vphantom{\dagger}} (x) \psi_p^{\dagger}(y) \colon -
\colon \psi_p^\dagger (x) \psi_p^{\vphantom{\dagger}}(y) \colon \right\}
\nonumber \\
+ s_p( x-y) s_p (x-z) \left\{
\colon \psi_p^{\vphantom{\dagger}} (y) \psi_p^\dagger(z) \colon -
\colon \psi_p^\dagger (y) \psi_p^{\vphantom{\dagger}}(z) \colon \right\}
\nonumber \\
- s_p( x-y) s_p (y-z) \left\{
\colon \psi_p^{\vphantom{\dagger}} (x) \psi_p^\dagger(z) \colon -
\colon \psi_p^\dagger (x) \psi_p^{\vphantom{\dagger}} (z) \colon \right\}.
\nonumber 
\end{eqnarray}
It coincides with the quadratic part of eq.(\ref{proof_:rrr:}). Therefore,
$\colon \rho_p^3 (x) \colon = G(x,x,x)$. We will demonstrate that
\begin{eqnarray}
G(x,x,x) = \ \colon\! \rho_p^3 (x) \colon =\label{Gx} \\
\frac{3}{4\pi^2} \left( \colon (\nabla \psi^\dagger_p(x) )
(\nabla \psi^{\vphantom{\dagger}}_p (x) ) \colon - \frac{1}{6} \nabla^2
\rho_p (x) \right). \nonumber 
\end{eqnarray}
We begin the proof by an observation:
\begin{eqnarray}
s_p (x-y) s_p (y-z) &=&s_p (x-y) s_p (x-z) + \\
&&s_p (y-z) s_p (x-z). \nonumber 
\end{eqnarray}
The formula can be checked directly with the help of eq.(\ref{s}).
We transform eq.(\ref{G}):
\begin{eqnarray}
G(x,y,z) = \qquad \\
s_p (x -y) s_p(x-z) \left\{ \colon \left[ \psi_p^\dagger (x) -
\psi_p^\dagger (y) \right]
 \psi^{\vphantom{\dagger}}_p (z) \colon + \text{h.c.} \right\} - \nonumber \\
s_p (y-z) s_p(x-z) \left\{ \colon \left[ \psi_p^\dagger (y) -
\psi_p^\dagger (z) \right]
 \psi^{\vphantom{\dagger}}_p (x) \colon + \text{h.c.} \right\}.  \nonumber
\end{eqnarray}
It is convenient to define another operator-valued function $F$:
\begin{eqnarray} 
G(x,y,z) =
s_p (x-z) \left[ F(x,y,z) - F(y,z,x) \right], \\
F(x,y,z) = \\
s_p(x-y) \left\{ \colon \left[ \psi_p^\dagger (x) - \psi_p^\dagger (y) \right] 
\psi^{\vphantom{\dagger}}_p (z) \colon + \text{h.c.} \right\}.  \nonumber 
\end{eqnarray}
Function is well-defined for any values of $x,y$ and $z$.

It is important to note that $F$ is invariant under the
transposition of its two first arguments:
\begin{eqnarray}
F(x,y,z) = \left( \delta (x-y) - s_p(y-x) \right) \times \\
\left\{ - \colon \left[ \psi_p^\dagger (y) - \psi_p^\dagger (x) \right] 
\psi^{\vphantom{\dagger}}_p (z) \colon + \text{h.c.} \right\} = F(y,x,z),
\nonumber
\end{eqnarray}
since $\delta(x-y) \colon\! [\psi^\dagger_p (x) - \psi^\dagger_p (y) ]
\psi^{\vphantom{\dagger}}_p (z) \colon = 0$.
Consequently, we can rewrite the expression for $G$:
\begin{eqnarray} 
G(x,y,z) = s_p (x-z) \left[ F(x,y,z) - F(z,y,x) \right].
\end{eqnarray}
The expression in the square brackets vanishes at $x=z$. Thus, function $G$
is well-defined everywhere. Sending $z \rightarrow x$ we obtain with the
help of the L'Hospital's rule:
\begin{eqnarray}
G(x,x,x) = \frac{i p}{2 \pi} \frac{\partial}{\partial z} \left[
F(x,x,z) - F(z,x,x) \right]\big|_{z=x}.
\end{eqnarray}
Applying the L'Hospital's rule the following equality is derived:
\begin{eqnarray}
F(x,x,z) = \frac{p}{2\pi i} \left\{ \colon (\nabla \psi_p^\dagger (x) )
\psi_p^{\vphantom{\dagger}} (z) \colon + \text{h.c} \right\}.
\end{eqnarray}
Consequently:
\begin{eqnarray}
\frac{\partial}{\partial z} F (x,x,z) \big|_{z=x} = 
\frac{p}{\pi i} \colon (\nabla \psi_p^\dagger (x) )
(\nabla \psi_p^{\vphantom{\dagger}} (x)) \colon.
\end{eqnarray}
We calculate:
\begin{eqnarray}
\frac{\partial}{\partial z} F (z,x,x) \big|_{z=x} = 
\frac{p}{4 \pi i} \left\{ \colon\! (\nabla^2 \psi^\dagger_p (x) )
\psi^{\vphantom{\dagger}}_p (x) \colon + \text{h.c.} \right\} \\
= \frac{p}{4 \pi i} \left\{\nabla^2 \rho_p(x) - 
2 \colon \! (\nabla \psi_p^\dagger (x) )
(\nabla \psi_p^{\vphantom{\dagger}} (x)) \colon \right\}, \nonumber 
\end{eqnarray}
by applying the L'Hospital's rule yet again.

Combining the results for $G$ and derivatives of $F$ we prove
eq.(\ref{Gx}) which, in turn, proves eq.(\ref{rho3}).

\end{document}